# An Android Cloud Storage Apps Forensic Taxonomy

M. Amine Chelihi, Akintunde Elutilo, Imran Ahmed, Christos Papadopoulos, Ali Dehghantanha

School of Computing, Science and Engineering, University of Salford, UK

amin.chelihi@hotmail.com, akinelutilo@yahoo.com, Imran@ebitm.co.uk, c.papadopoulos@salford.ac.uk, A.Dehghantanha@salford.ac.uk

## ABSTRACT

Mobile phones have been playing a very significant role in our daily activities for the last decade. With the increase need for these devices, people are now more reliant on their smartphone applications for their daily tasks and many prefer to save their mobile data on a cloud platform to access them anywhere on any device. Cloud technology is the new way for better data storage, as it offers better security, more flexibility, and mobility. Many smartphones have been investigated as subjects, objects or tools of the crime. Many of these investigations include analysing data stored through cloud storage apps which contributes to importance of cloud apps forensics on mobile devices. In this paper, various cloud Android applications are analysed using the forensics tool XRY and a forensics taxonomy for investigation of these apps is suggested. The proposed taxonomy reflects residual artefacts retrievable from 31 different cloud applications. It is expected that the proposed taxonomy and the forensic findings in this paper will assist future forensic investigations involving cloud based storage applications.

**INTRODUCTION**

Mobile applications have increasingly adopted the Android operating system as a standard, with backing from market share figures, 82.8% of smartphones have incorporated the Android operating system (Hornyak, 2014). These statistics and features highlight the importance and relevance of the Android OS for smartphone manufacturers worldwide (Kumar & Xie, 2012). Android operating system has amassed over 1.4 billion users worldwide as of 2016 according to the Google CEO (Vincent, 2015) this is due to its wide acceptability and increased usage amongst the world's mobile phone populace (Stevenson, 2014). Cloud computing is an umbrella under which majority of internet related services are classified. Many cloud applications are increasingly developed for Android platform and many Android applications are now running through cloud services (Tee, 2012). Therefore, more attacks are directed against Android platforms through cloud services or cloud applications vulnerabilities (Ganji, Dehghantanha, & IzuraUdzir, 2013a). As the use of technology has afforded substantial breakthroughs in investigative cases, it is now required that the forensic investigators be able to acquire and analyse persistent data, volatile data and other sensitive information from different platforms (Damshenas & Dehghantanha, 2014). Many investigation tools were developed to assist acquisition, collation, and analysis of data in ways that maintain the chain of evidence (Massie, 2014). Mobile forensic investigators are required to integrate technology and law in investigation of many traditional crimes such as drug dealing, copyright infringement, etc, which are facilitated by mobile and cloud technologies (Mohsen Damshenas, Ali Dehghantanha, Ramlan Mahmoud, 2013; Yusoff, Mahmod, & Abdullah, 2014). Audio files, images, videos, messages, notes and emails are just a few possible remnants that can be extracted during mobile investigations from social networking, instant messaging and cloud and mobile platforms (Aminnezhad & Dehghantanha, 2013; Daryabar et al., 2016; Seyedhossein Mohtasebi & Dehghantanha, 2011; SH Mohtasebi, 2011; Yang, Dehghantanha, Choo, & Muda, 2016). As the cyber world becomes heavily reliant on mobile and cloud storage technologies, it is only palpable that these two mediums would experience the most cyber-attacks and intrusion attempts as they hold vital information of the concerned organizations or individuals (Farid Daryabar, Ali Dehghantanha, Nur Izura Udzir, Nor Fazlida binti Mohd Sani, 2013).

Methods, techniques and strategies must be implemented to not only defend or protect these mediums but also to ensure their forensics readiness (Mohsen Damshenas, Ali Dehghantanha, Ramlan Mahmoud, 2013). The fast pace of change in mobile and cloud technologies mandates utilising many different tools and techniques for investigation of these platforms (Daryabar & Ali Dehghantanha, Nur Izura Udzir, Nor Fazlida binti Mohd Sani, 2013). Previous researchers (A Azfar, Choo, & Liu, 2016; Abdullah Azfar, Choo, & Liu, 2015a, 2015b) have carried out experiment on applications to determine what user information or activity the average applications collects with a view to improving the forensic understanding of the tested applications by creating taxonomies from their studies.

The taxonomy created in this paper would aid in investigation of cloud based storage applications and reflect residual artefacts of 31 different cloud apps on an android mobile device which aids in correlation of evidences between user's activities and remnants of cloud applications resided on the device.

The remainder of this paper is structured as follows, firstly, the experiment set-up which details the cloud storage applications, environment, devices and operating systems used. Secondly, results and discussion which illustrates our findings and end results of this experiment. Finally, the paper is concluded and several future researches are suggested.

## EXPERIMENT SETUP

In order to select cloud storage application that would cover majority of users, two major criteria were considered, targeting free applications and also the user ratings for the cloud storage applications. This gives basis for strong and efficient cloud storage application taxonomy to be developed. Free applications were selected as majority of users would rather download the free cloud storage application as they do not require commercial services. In return of search query "cloud storage applications" 240 free Cloud storage applications were shortlisted and only 31 were selected from these based on their user ratings and download numbers(Falk & Shyshka, 2014a). 31 applications which best fit the selection criterial were chosen. 31 cloud storage applications are deemed adequate in accordance with previous works (Yang et al., 2016) in the forensic investigation of cloud storage services.

The selected applications can be found in appendix 1.In this paper, Android cloud storage applications were examined using the popular forensics tool MicroSystemation XRY (version 6.16.0) from MSAB on a Windows operating system (Win 10 Pro). Cloud storage applications were sourced from Google Play Store. Applications selected are all available and free to download. The applications were picked out from a pool of over 240 cloud applications (the total number of cloud applications on google play as at 10/03/2016) however the scope of this project only focuses on cloud-storage based applications.

A sheer number of users utilise those popular apps daily, analysing those apps would cover a high percentage of possible cases that may require forensics investigation, there might be a good chance to examine vulnerabilities that could be exploited by cyber criminals under any circumstances. When investigating mobile devices that are likely to contain data which could be decisive in criminal investigations or which can be presented as forensic evidence, it is expected that files and applications that have been executed or transferred on the mobile devices can still be accessed if the need arises. Files and data on devices can still be retrieved from mobile device even after they have been deleted. The data extraction tool has the capability to pull out data that has been removed on various applications depending on the security and technological level of the application.

XRY (version 6.16.0) is installed on the machine in order to carry out the forensic extraction and analysis legally. The Asus Nexus 7 Google tablet was used for the running and analysis of the selected android cloud storage applications. The analysis was conducted on Android operating system version KitKat (4.4.4) due to the difficulty the XRY tool encountered trying to capture after upgrading Asus Nexus 7 Google tablet to the latest Android version Marshmallow (6.0.1). A version downgrade was required in order to resolve this issue, the downgrade was done using Nexus root Toolkit (2.4.1) (Nammi, 2014).

Factory reset was carried out after examining each cloud storage application, this was done to ensure that no interferences or mix up from previously examined applications was recorded as part of the result for the subsequent application examined. This technique also eliminated the possibility of any dataset remnants from the previous examined application will not be carried over to subsequent data extractions thereby guaranteeing a more accurate investigation.

The Nexus Root Toolkit (2.4.1) was used for the rooting of the Asus Nexus 7 Google tablet, this toolkit is widely used for rooting Nexus devices and furthermore as XRY was unable to automatically root the device (Mushcab & Gladyshev, 2015). Operating system vendors have restriction on the privileges, access and rights that normal users do not have the authentication to access so in order to gain maximum privileges and full control of the Android operating system (Android 4.4.4 (KitKat)). This would allow access with root privileges, applications on the device can now run with privileged commands hereby giving access to control the CPU and Kernel of the device (Swift, 2015).

In order to sniff only network traffic generated by the examined applications and not by other services and programs running in the background, an separated hotspot was created in order to avoid interference from devices or applications on the same network. Similar to authors of previous works (Park, Choi, Eom, & Chung, 2014) in creating a hotspot in order to meet this very important pre requisite. The Asus Nexus 7 Google tablet was placed on a separate network established using Connectify (2016.0.3.36821 Pro) which allowed having a hotspot with specific IP addresses and this separated the Asus Nexus 7 Google tablet’s IP address from the local network. It must be exclusively network traffic between the applications and their communications between the Asus Nexus 7 Google tablet, and the server on an isolated network.

As done in previous researches (Ashraf, 2013; Lo, Qian, Chen, & Rogers, 2015) Connectify was used to set up a hotspot connection, which connects the tablet to the laptop and separates it from the main local network, the laptop then works as a server or a gateway for the Asus Nexus 7 Google tablet and hence making it easier to monitor the traffic sent from the Asus Nexus 7 Google tablet to the server. Wireshark (2.0.2) was used to monitor and capture the traffic of the Asus Nexus 7 Google tablet on that specific connection. The IP address of the Asus Nexus 7 Google tablet is fixed (192.168.121.103) as used for this project. This IP address would serve as the source while the IP address of the server is (192.168.121.1) as it works as a gateway for the Asus Nexus 7 Google tablet. Wireshark (2.0.2) has also been used by (Ashraf, 2013) for sniffing network traffic in similar forensic work.

For appraisal of data extracted by the XRY tool for this experiment, a forensic dataset was chosen and files were selected based on certain criteria and utilized for the purpose of the experiment. The chosen files were adopted as exhibits that would be uploaded for every cloud storage application. This gives a fair and accurate ground that the files and documents used are of optimal integrity and were compromised to alter results. We chose the EDRM dataset which provides industry standard reference dataset of electronically stored information (ESI) used for forensic and other e-discovery works (EDRM, 2010). Specifically 11 files of different formats and extension, the files chosen were based on the need to cover the most common file extensions and types which cater to different media and document contents. In line with works by authors Dehghantanha, A Daryabar, F. et al. (Daryabar et al., 2016) who made use of the EDRM dataset in their forensic research on similar cloud storage applications. The dataset is exhibited in the appendix 2 of this paper.

## DISCUSSION

Extracted data from each application were examined and the datasets were identified, presented and documented. These traces, effects and remnants of the files together with other activities or action by the cloud storage application on the Asus Nexus 7 Google tablet were tabulated.

The reason behind choosing the corresponding multimedia files for this experiment, is due to the fact that forensics investigators usually tend to capture multimedia remnants from the suspect's device, this will most probably be used as a proof of evidence in the court, files like images and audio files contain a lot of sensitive data, one image could simply open or close the case. Documents are of a high importance too, in the real world, many leaked documents contained a sheer amount of sensitive data, and that is due to lack of proper security for these documents. Having put documents of certain information in this experiment, helped us track that information while searching for the documents' remnants among the extracted data. Often when the documents are deleted, they leave behind some remnants which sometimes can be in a readable format, even though the data may not be complete, it sometimes gives an indication on what the full information was. Supposedly, in a real life's forensics investigation process, a suspect's device is used to store a couple of documents on the cloud, any retrieved information from the document file may hold a high amount of valuable information that could be presented as a proof of evidence.

Internal storage contains installed applications and their data are stored on mobile devices and sometimes they leave some remnants, this information can be potentially sensitive or incriminating therefore they play a pivotal role in the forensics investigation of suspected mobile devices in cases of law prosecution and forensic investigations of such devices.

Internal memory though volatile in nature, is much faster to write to and read from in comparison with other forms of storage such as hard disk drives. Data in the internal memory normally remains intact while the device is powered on but loses it when the power goes off (Microsoft, 2006). This previously hindered forensic investigation of such part of the memory but with the advent of tools such as XRY it is now possible to recover some of this data through acquisition. This is also shown in Appendix 3 in the remnants column (those remnants were found in the internal storage as cache files)

Network analysis is important to monitor, capture and analyse network traffic packets which hold network information. The network analysis provides investigators with relevant IP addresses, ports, protocols and URL's assessed by the cloud storage applications as well as attempted connections (Shimonski, 2013). This can prove to be vital information in tracing criminal activities. Appendix 4 depicts some of the network traffic observed when these applications were sharing files on the internet.

The protocols used by these applications were identified and listed with check marks denoting which application makes use of which protocols; these can be crucial for forensic investigating and ethical hacking of devices and networks. The UDP stream from the DNS and MDNS show paths taken by some application when users upload content unto the cloud via their application, these paths can be used varying and specifically based on the motive for investigation, analysis or monitoring network traffic. The streams give a little insight into the kinds of information that these applications generate as they peruse the networks and store user content on the cloud platforms. As shown in Appendix 4 some of the applications display the paths they take during installation, registration and running of services. Few applications leave traces in plain text of their origin website, which is likely to be the provider which hosts these cloud services.

## RESULTS

The applications examined share a similar criteria, each application stores data on the cloud so it could be accessible by the user either from the phone (device) or using a website (in most cases). Some applications selected use a third party cloud service in order to store data on the cloud, which may come in handy especially when the app offers other services like managing files or exploring the phone's storage.

Typically, each application requires a sign up in order to use its cloud service, and for most applications, we were able to upload data on the cloud from their home website.

For better results, we ensured that the dataset was uploaded to the cloud service using a different device other than the tablet, to avoid having remnants from the internal storage interfering with the data retrieved from the application itself, even though XRY lists out the data found and the path of where the data was found particularly. Remnants found on */mnt/sdcard/* are discarded, for some applications the dataset sample was transferred directly to the tablet, and then uploaded to the cloud application, which is why some remnants could be recovered from that particular path and not from the app itself.

### Database

When extracting the data, we noticed that some applications generated *db* files and stored them in a particular path within the internal storage, among the experimented applications, 15 apps generated *db* files in the internal memory, and the following is a list of the apps along with the paths where the *db* files were stored:

- *File Expert* */data/data/xcxin.fehd/databases/*
- *GCloud* */data/data/com.genie9.gcloudbackup/databases/*
- *Mail.Ru* */data/data/ru.mail.cloud/databases/*
- *Cloud Magic* */data/data/com.cloudmagic.mail/databases/*
- *Degoo* */data/data/com.degoo.android/databases/*
- *Pcloud* */data/data/com.pcloud.pcloud/databases/*
- *XXL Box* */data/data/com.xxlcloud.xxlbox/databases/*
- *ZeroPC* */data/data/com.zeropc.tablet/databases/*
- *File Manager* */data/data/fm.clean/databases/*
- *4Sync Cloud* */data/data/com.forsync/databases/*
- *Asus Web* */data/data/com.ecareme.asuswebstorage/databases/*
- *FileBay.co* */data/data/com.sflcnetwork.filebayco/databases/*
- *Folder Sync* */data/data/dk.tacit.android.foldersync.lite/databases/*
- *MyCloud* */data/data/com.wdc.wd2go/databases/*
- *Pogoplug* */data/data/com.wdc.wd2go/databases/*

The applications intended here are under both groups 2 and 3, as applications from Group 1 did not generate any db files.

Notice that the path for this particular file, always shows same directory names: "*data/data/generated_file_name/databases/"* this kind of paths is inaccessible to the user. Check Appendix3 for more information on the applications and their db files.

**Storage**

The artefacts retrieved from the dataset are usually found on the internal storage folder which is accessible by the user, such as the pictures folder, the music folder and so on. A good observation would be that some images were not retrieved and yet their thumbnails were, and from a simple thumbnail a lot of information may be obtained during the forensics investigation process, it may not show all the details as the number of pixels in thumbnails is obviously much lower than the those of an original image, however, using some image processing tools, a lot of evidence could be extracted from a simple thumbnail.

The following lists the findings of the extracted data within the internal storage, which includes **Pictures, Documents, Audio files,** and **Web files**.

- **Pictures**

We were able to retrieve images from some applications, as mentioned before, those retrieved images may not have the same size, resolution and file details as the original ones, however, those images were viewable, and are enough to be shown as evidence for forensics investigation purposes. The following applications showed recovered artefacts (i.e. original pictures, thumbnails, cache images):

- *Box*
- *GCloud*
- *Mail.Ru*
- *Cloud Magic*
- *Just Cloud*
- *XXL Box Secure*
- *Zero PC*
- *File Manager*
- *ZipCloud*
- *Folder Sync Lite*
- *MyCloud*

See **Appendix 3** for the full list of applications and their respective findings.

- **Documents**

For some applications, the following documents were retrieved from the internal storage and were in a readable format too: *.doc, .pot and .pdf*. Most apps did not show any retrieved document files during the extraction, except for a minority of apps, those are as follow:

- *Cloud Magic*
- *XXL Box Secure*
- *File Manager*
- *DropSend Android Cloud*
- *Folder Sync Lite*
- *MyCloud*

- **Web Files**

A variety of applications stored web files from the correspondent dataset used in the experiment (*.xml and .html*) in the internal storage and hence it was retrievable and readable too. The apps are as follow:

- *XXL Box Secure*
- *File Manager*
- *Zip Cloud*
- *DropSend Android*

- *Folder Sync Lite*
- *MyCloud*

- **Audio**

The audio file from the selected dataset was recovered from a minority of applications only and was playable, those applications are:

- *Just Cloud*
- *XXL Box Secure*
- *File Manager*
- *Folder Sync Lite*
- *My Cloud*

Upon analysing the extracted data, the path of the retrieved audio file showed the following: "*mnt/sdcard"* that is an evidence of the existence of data or artefact on the internal storage of the device, and is accessible by the user. While the path "*/data/data/"* may not be accessible by a regular user, but that is not the case for any forensic investigator examining the device for evidence. **Appendix 3** illustrates these results and clarifies the retrieved files for each of the 31 applications.

As a conclusion to these results, we can state that based on the amount of files retrieved from each app in this experiment we could categorise them into 3 groups according to how much data (artefact) was recovered, the first group of applications showed no recovered data on XRY, which we will classify in this paper as (**Group 1**), while some applications generated database files in the internal storage only while the files from the dataset were not recovered, we call this group (**Group 2**), and finally (**Group 3**) showed database files as well as most of the data stored in the cloud was retrieved. As shown in **Table 1**, cloud applications used in this experiment can be categorised into 3 groups.

| **Applications Classification** | | |
|---|---|---|
| **Group 1** | **Group 2** | **Group 3** |
| Box | File Expert | Mail.Ru |
| DropBox | Degoo | GCloud |
| Google Drive | PCloud | Cloud Magic |
| Adobe Creative Cloud | 4Sync Cloud | Just Cloud |
| One Drive | Asus Web Storage Cloud | XXL Box Secure |
| Yandex | FileBay.co | ZeroPC |
| Tesorit | PogoPlug Cloud | File Manager |
| Egnyte | | Zip Cloud |
| Bitcasa | | Drop Send Android Cloud |
| IDrive Cloud | | Folder Sync Lite |
| JohoSpace | | My Cloud (WD) |
| Mega Cloud | | |

*Table 1: Classification of apps based on this study.*

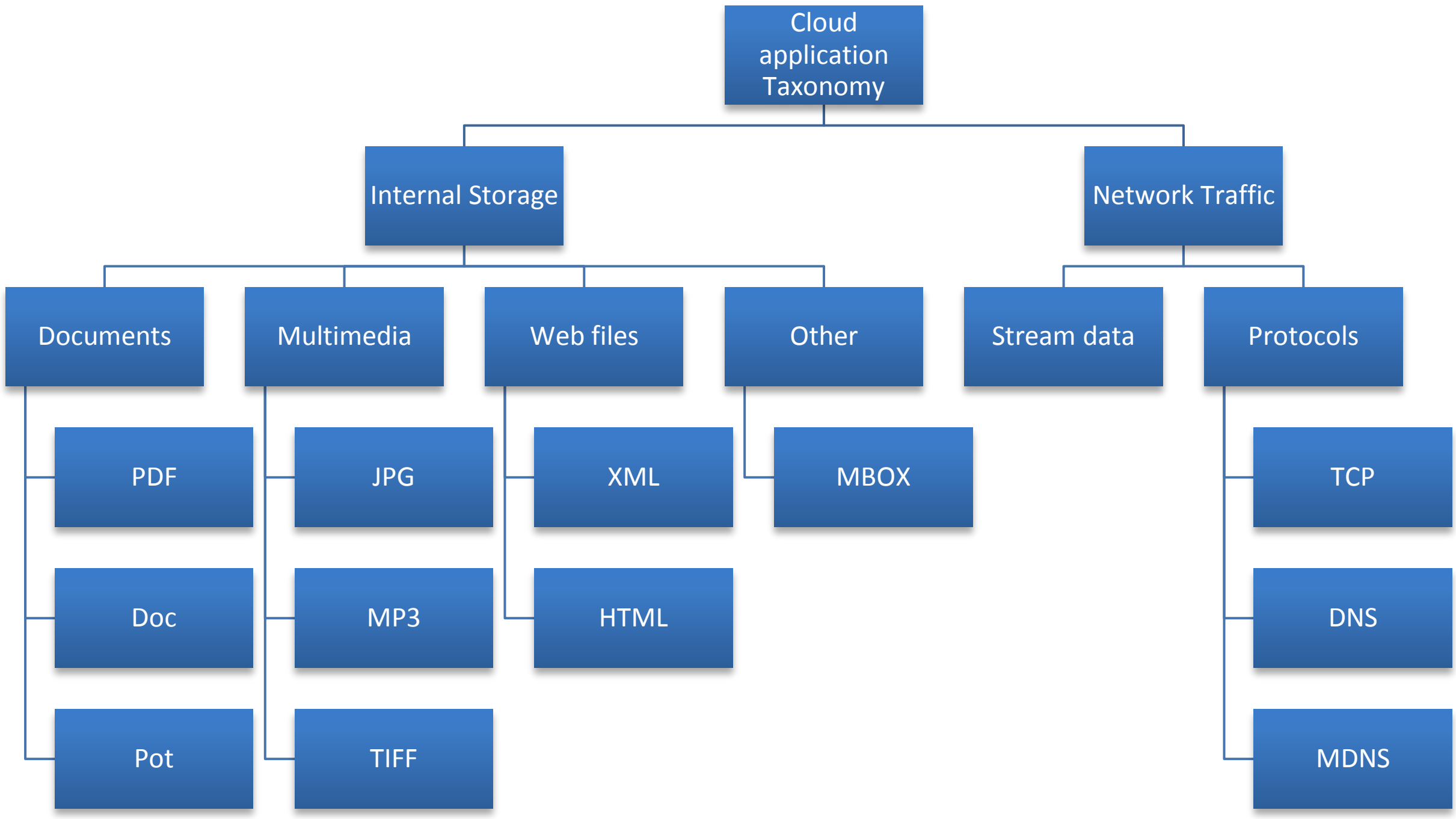

*Figure 1: Classification of data in this experiment*

The diagram in Figure 1 illustrates the classification of data recovered during this taxonomy forensics experiment, some applications showed a lot of recovered data, while others showed no trace of those files, this diagram facilitates an understanding on how to categorise and tabulate the data recovered. This Taxonomy was coined from the intricate analysing of the recovered remnant from the extraction processes. It is a comprehensive and relevant taxonomy in the field of cyber cloud forensics. This is depicted in the detailed results and discussions sections of this paper.

## CONCLUSION & FUTURE WORKS

Mobile Forensics, and specifically Android Forensics, continues to be a growing field of research. As the Android platform and its architectures continue to evolve, the current research will facilitate forensic investigators in recovering sources of evidence (Martini, Do, & Choo, 2015a, 2015b; Shariati, Dehghantanha, Martini, & Choo, 2015).

The main purpose of this paper is to contribute to the forensics field and particularly the Android mobile forensics. With a taxonomy experiment on Android cloud applications, we deem its necessity in this particular area of research as the taxonomy for cloud applications is minimal. Using the taxonomy and forensics model, we conducted a fundamental analysis on 31 popular cloud applications; the findings are tabulated and presented appropriately in the Appendix section.

Finally, as part of our future work, we intend to examine other cloud applications on other platforms (i.e. IOS) and devices, and push the investigation by performing physical data acquisition on those apps in order to capture dumps from the device's memory which may lead to even more promising findings. However, more research is still needed in order to provide better directions on addressing different challenges of mobile forensics, and prompt further discussion on the development of different forensic taxonomies in variety of contexts.

**BIBLIOGRAPHY**

Aminnezhad, A., & Dehghantanha, A. (2013). Cloud Forensics Issues and Opportunities . *InternationalJournal of Informaytion Processing and Management*, *4*(4), 76. Retrieved from https://www.researchgate.net/profile/Mohd_Taufik_Abdullah2/publication/276010807_Cloud_Forensics_Issues_and_Opportunities/links/5556f1cb08ae980ca60c9daf.pdf

Ashraf, R. A. (2013). *Performance Analysis Of Video Call Application On Tablet Using 3G Network*. Retrieved from http://eprints.utem.edu.my/13869/

Azfar, A., Choo, K. R., & Liu, L. (2015a). Forensic Taxonomy of Popular Android mHealth Apps. In *Proceedings of the 21st Americas Conference on Information Systems (AMCIS 2015)*. Retrieved from http://aisel.aisnet.org/cgi/viewcontent.cgi?article=1217&context=amcis2015

Azfar, A., Choo, K.-K. R., & Liu, L. (2015b). An Android Communication App Forensic Taxonomy. *Journal of Forensic Sciences*.

Azfar, A., Choo, K.-K. R., & Liu, L. (2016). Forensic Taxonomy of Android Social Apps. *Journal of Forensic Sciences*.

Damshenas, M., & Dehghantanha, A. (2014). A survey on digital forensics trends. *International Journal of Cyber-Security and Digital Forensics*, *3*(4), 235. Retrieved from http://go.galegroup.com/ps/i.do?id=GALE%7CA387349458&sid=googleScholar&v=2.1&it=r&linkaccess=fulltext&issn=23050012&p=AONE&sw=w

Daryabar, F., & Ali Dehghantanha, Nur Izura Udzir, Nor Fazlida binti Mohd Sani, S. bin S. (2013). A REVIEW ON IMPACTS OF CLOUD COMPUTING ON DIGITAL FORENSICS. *International Journal of Cyber-Security and Digital Forensics (IJCSDF)*, *2*(2), 77–94.

Daryabar, F., Dehghantanha, A., Eterovic-Soric, B., Choo, K.-K. R., Federici, C., Daryabar, F., … Eterovic-Soric, B. (2016). Forensic investigation of OneDrive, Box, GoogleDrive and Dropbox applications on Android and iOS devices. In *Australian Journal of Forensic Sciences* (Vol. 48, pp. 1–28). http://doi.org/10.1080/00450618.2015.1110620

Dehghantanha, A. (n.d.-a). Cloud and Embedded Forensics Extracting footprints from cloud storage platforms and analyzing embedded computing devices. Retrieved from http://www.alid.info/cloud-forensics

Dehghantanha, A. (n.d.-b). Mobile Phone Forensics Extracting evidences from locked or even damaged smart phones. Retrieved from http://www.alid.info/mobile-phone-forensics

EDRM. (2010). New EDRM resource provides emerging data types for e-discovery testing New EDRM Enron Email Data Set. Retrieved from http://www.edrm.net/resources/data-sets/edrm-enron-email-data-set

Falk, S., & Shyshka, A. (2014a). The Cloud Marketplace: A Capability-Based Framework for Cloud Ecosystem Governance. Retrieved from http://www.diva-portal.org/smash/record.jsf?pid=diva2:721103

Falk, S., & Shyshka, A. (2014b). *The Cloud Marketplace: A Capability-Based Framework for Cloud Ecosystem Governance*. Retrieved from http://www.diva-portal.org/smash/get/diva2:721103/FULLTEXT01.pdf

Farid Daryabar, Ali Dehghantanha, Nur Izura Udzir, Nor Fazlida binti Mohd Sani, S. B. S. (2013). A Survey about Impacts of Cloud Computing on Digital Forensics. *International Journal of Cyber-Security and Digital Forensics*, *2*(2), 77–94. Retrieved from http://sdiwc.net/security-journal/Browse-Archive.php?ptid=1&ptsid=66&vnum=2&inum=2

Ganji, M., Dehghantanha, A., & IzuraUdzir, N. (2013a). Cyber warfare trends and future. *Advances in Information Sciences and Service Sciences*, *5*(13), 1. Retrieved from http://search.proquest.com/openview/cdd017512e93963c0f2ac8f2e88d3c5f/1?pq-origsite=gscholar

Ganji, M., Dehghantanha, A., & IzuraUdzir, N. (2013b). Cyber warfare trends and future. *Advances in*. Retrieved from http://search.proquest.com/openview/cdd017512e93963c0f2ac8f2e88d3c5f/1?pq-origsite=gscholar

Hornyak, T. (2014). Android grabs record 85 percent smartphone share. Retrieved February 18, 2016, from http://www.pcworld.com/article/2460020/android-grabs-record-85-percent-smartphone-share.html

Kumar, A., & Xie, B. (2012). *Handbook of Mobile Systems Applications and Services*. CRC press . Retrieved from https://books.google.co.uk/books?hl=en&lr=&id=VCWMCgAAQBAJ&oi=fnd&pg=PP1&dq=2.%09Kumar,+A.+%26+Xie,+B.+(2012).+Handbook+of+mobile+systems+applications+and+services.+Boca+Raton:+CRC+Press&ots=iM7-3IZaHM&sig=wv6RShQVPBywH_wtUzJbtg9D8BI

Lo, D. C. T., Qian, K., Chen, W., & Rogers, T. (2015). A low cost, portable platform for information assurance and security education. In *Proceedings - IEEE 15th International Conference on Advanced Learning Technologies: Advanced Technologies for Supporting Open Access to Formal and Informal Learning, ICALT 2015* (pp. 111–113). http://doi.org/10.1109/ICALT.2015.132

Martini, B., Do, Q., & Choo, K.-K. R. (2015a). Chapter 15 - Mobile cloud forensics: An analysis of seven popular Android apps. In *The Cloud Security Ecosystem* (pp. 309–345). http://doi.org/http://dx.doi.org/10.1016/B978-0-12-801595-7.00015-X

Martini, B., Do, Q., & Choo, K.-K. R. (2015b). Conceptual evidence collection and analysis methodology for Android devices. In *The Cloud Security Ecosystem* (pp. 285–307). http://doi.org/10.1016/B978-0-12-801595-7.00014-8

Massie, R. (2014). What is Cyber ? Retrieved from https://www.isc2.org/cyber-forensics.aspx

Microsoft. (2006). Volatile and Nonvolatile Storage Devices. Retrieved from https://msdn.microsoft.com/en-us/library/ms940147(v=winembedded.5).aspx

Mohsen Damshenas, Ali Dehghantanha, Ramlan Mahmoud, S. bin S. (2013). Cloud Computing and Conflicts with Digital Forensic Investigation. *International Journal of Digital Content Technology and Its Applications*, *7*(9), 543. Retrieved from

http://search.proquest.com/openview/43636b97fc3d6f8d5332b32bb9c1637e/1?pq-origsite=gscholar

Mohtasebi, S. (2011). Smartphone Forensics: A Case Study with Nokia E5-00 Mobile Phone. *Journal of Digital Forensics, Security and Law*, *1*(3), 651–655. Retrieved from http://sdiwc.net/digital-library/smartphone-forensics-a-case-study-with-nokia-e500-mobile-phone

Mohtasebi, S., & Dehghantanha, A. (2011). Defusing the Hazards of Social Network Services. *International Journal of Digital Information and Wireless Communications (IJDIWC)*, *1*(2), 504–516. Retrieved from http://sdiwc.net/digital-library/defusing-the-hazards-of-social-network-services.html

Mushcab, R. Al, & Gladyshev, P. (2015). The significance of different backup applications in retrieving social networking forensic artifacts from Android-based mobile devices. In *Second International Conference on Information Securtiy and Cyber Forensics* (pp. 66–71). Retrieved from http://ieeexplore.ieee.org/xpls/abs_all.jsp?arnumber=7435508

Nammi, S. (2014). How to revert back from Android L to KitKat. Retrieved from http://nexusandme.com/revert-back-from-android-l-to-kitkat/

Park, M. W., Choi, Y. H., Eom, J. H., & Chung, T. M. (2014). Dangerous Wi-Fi access point: Attacks to benign smartphone applications. *Personal and Ubiquitous Computing*, *18*(6), 1373–1386. http://doi.org/10.1007/s00779-013-0739-y

Shariati, M., Dehghantanha, A., Martini, B., & Choo, K.-K. R. (2015). Chapter 19 - Ubuntu One investigation: Detecting evidences on client machines. In *The Cloud Security Ecosystem* (pp. 429–446). http://doi.org/http://dx.doi.org/10.1016/B978-0-12-801595-7.00019-7

Shimonski, R. (2013). *The Wireshark Field Guide: Analyzing and Troubleshooting Network Traffic*. *The Wireshark Field Guide*. Retrieved from http://www.sciencedirect.com/science/article/pii/B9780124104136000012

Stevenson, A. (2014). Top 10 Android benefits over Apple iPhone. Retrieved from http://www.v3.co.uk/v3-uk/news/2336474/top-10-android-benefits-over-apple-iphone/page/3?mt=143&mv2=370&mv=969

Swift, O. (2015). *Android App Development & Programming Guide*.

Tee, J. (2012). Top five ways cloud computing impacts mobile application development teams. *TheServerSide.com*. Retrieved from http://www.theserverside.com/feature/Top-five-ways-cloud-computing-impacts-mobile-application-development-teams

Vincent, J. (2015). Android is now used by 1.4 billion people. Retrieved from http://www.theverge.com/2015/9/29/9409071/google-android-stats-users-downloads-sales

Yang, T. Y., Dehghantanha, A., Choo, K.-K. R., & Muda, Z. (2016). Windows Instant Messaging App Forensics: Facebook and Skype as Case Studies. *PloS One*, *11*(3), e0150300. http://doi.org/10.1371/journal.pone.0150300

Yusoff, M., Mahmod, R., & Abdullah, M. (2014). Mobile forensic data acquisition in Firefox OS. *Cyber Security, Cyber Warfare and Digital Forensics*, *3*(4), 209–235. Retrieved

## APPENDIX

### Appendix 1 – List of Application versions.

| App ID. | App Name | App Version |
|---|---|---|
| 1. | *Box* | 3.7.5 |
| 2. | *File Expert* | 7.1.8 |
| 3. | *GCloud* | 5.4.95 |
| 4. | *Mail.Ru* | 4.4.0.15562 |
| 5. | *MEO Cloud* | 2.1.0 |
| 6. | *Cloud Magic* | 8.1.32 |
| 7. | *Drop Box* | 5.2.6 |
| 8. | *Degoo* | 1.9.7 |
| 9. | *One Drive* | 3.9.1 |
| 10. | *Just Cloud* | 1.2.4 |
| 11. | *PCloud* | 1.12.05 |
| 12. | *XXL Box Secure* | 1.0.2 |
| 13. | *ZeroPC* | 2.5.5 |
| 14. | *File Manager* | 4.0.4.9 |
| 15. | *ZipCloud* | 1.2.4 |
| 16. | *4Sync Cloud* | 3.27.0 |
| 17. | *Adobe Creative Cloud* | 2.3.460 |
| 18. | *Asus Web Storage Cloud* | 3.1.8.0408 |
| 19. | *Bitcasa Drive* | 4.25.5 |
| 20. | *DropSend Android Cloud* | 1.0.0-rc6 |
| 21. | *Egnyte* | 6.8.5 |
| 22. | *FileBay.co Secure Storage* | 1.0 |
| 23. | *Foldersync Lite* | 2.8.5.90 |
| 24. | *IDrive Cloud Online Backup* | 3.6.32 |
| 25. | *JohoSpace Backup – Restore & Migrate* | 3.6.3 |
| 26. | *Mega Cloud* | 3.0.6.official |
| 27. | *MyCLoud (WD)* | 4.4.3 |
| 28. | *Pogoplug Cloud* | 5.11.0.19 |
| 29. | *Tesorit Cloud* | 2.1.386.443 |
| 30. | *Yandex Disk* | 2.65 |
| 31. | *Google Drive* | 2.4.141.16.34 |

**Appendix 2 – Dataset List (EDRM)**

| Dataset ID | File Name and Extension | File Type |
|---|---|---|
| 1. | *Student-Documentation.pdf* | PDF Document |
| 2. | *Malaventura_-_12_The_Queen.mp3* | Media File Audio |
| 3. | *IMG_3084.jpg* | JPEG Image |
| 4. | *IMG_3085.jpg* | JPEG Image |
| 5. | *EDRM.html* | HTML Document |
| 6. | *Disposing-of-Digital-Debris-Information-Governance-Practice-and-Strategy_Page_02.tif* | TIFF Image |
| 7. | *Disposing-of-Digital-Debris-Information-Governance-Practice-and-Strategy_Page_05.tif* | TIFF Image |
| 8. | *Disposing of Digital Debris Sample 1.pot* | Microsoft PowerPoint Document |
| 9. | *Disposing of Digital Debris – 97.doc* | Microsoft Word Document |
| 10. | *DERM Statistical Data Sample 1 Unicode.xml* | XML Document |
| 11. | *Cindyloh3333@gmail.com.mbox* | MBOX e-mail file |

**Appendix 3 – Retrieved Artefacts**

| App ID | App name | DB Files | Path | JPEG | TIF | DOC | POT | PDF | XML | HTML | Mp3 | MBOX |
|---|---|---|---|---|---|---|---|---|---|---|---|---|
| 1. | Box | N/A | N/A | ✓ | ✗ | ✗ | ✗ | ✗ | ✗ | ✗ | ✗ | ✗ |
| 2. | File Expert | FileExpert.db | /data/data/xcxin.fehd/databases/ | ✗ | ✗ | ✗ | ✗ | ✗ | ✗ | ✗ | ✗ | ✗ |
| 3. | GCloud | GCloudDB, db_music, db_video, syncdata.db | /data/data/com.genie9.gcloudbackup/databases/ | ✓ | ✗ | ✗ | ✗ | ✗ | ✗ | ✗ | ✗ | ✗ |
| 4. | Mail.Ru | adman.db, isp.taxonomy@mail.ru, google_analytics_v4.db | /data/data/ru.mail.cloud/databases/ | ✓ | ✓ | ✗ | ✗ | ✗ | ✗ | ✗ | ✗ | ✗ |
| 5. | MEO Cloud | N/A | N/A | ✗ | ✗ | ✗ | ✗ | ✗ | ✗ | ✗ | ✗ | ✗ |
| 6. | Cloud Magic | google_analytics_v2.db | /data/data/com.cloudmagic.mail/databases/ | ✓ | ✗ | ✓ | ✗ | ✓ | ✗ | ✗ | ✗ | ✗ |
| 7. | **Drop Box** | N/A | N/A | ✗ | ✗ | ✗ | ✗ | ✗ | ✗ | ✗ | ✗ | ✗ |
| 8. | Degoo | google_analytics_v4.db | /data/data/com.degoo.android/databases/ | ✗ | ✗ | ✗ | ✗ | ✗ | ✗ | ✗ | ✗ | ✗ |
| 9. | **One Drive** | N/A | N/A | ✗ | ✗ | ✗ | ✗ | ✗ | ✗ | ✗ | ✗ | ✗ |
| 10. | Just Cloud | N/A | N/A | ✓ | ✗ | ✗ | ✗ | ✗ | ✗ | ✗ | ✓ | ✗ |
| 11. | PCloud | PCloudDB, PCloudDB_wal, PCloudDB_shm, google_analytics_v4.db | /data/data/com.pcloud.pcloud/databases/ | ✗ | ✗ | ✗ | ✗ | ✗ | ✗ | ✗ | ✗ | ✗ |
| **12.** | XXL Box Secure | Monitor.db, data.db, account.db | /data/data/com.xxlcloud.xxlbox/databases/ | ✓ | ✓ | ✓ | ✓ | ✓ | ✓ | ✓ | ✓ | ✓ |
| **13.** | ZeroPC | com.zeropc.tablet.queue.db, com.zeropc.tablet.cache.db | /data/data/com.zeropc.tablet/databases/ | ✓ | ✗ | ✗ | ✗ | ✗ | ✗ | ✗ | ✗ | ✗ |
| **14.** | File Manager | clanfilemanager.db, google_tagmanager.db, google_analytics_v4.db | /data/data/fm.clean/databases/ | ✓ | ✓ | ✓ | ✓ | ✓ | ✓ | ✓ | ✓ | ✓ |
| **15.** | ZipCloud | N/A | N/A | ✓ | ✗ | ✗ | ✗ | ✗ | ✓ | ✗ | ✗ | ✓ |
| **16.** | 4Sync Cloud | Cloud.db, sdk4.uploads.db, sdk4.uploads.db-shm, sdk4.uploads.db-wal, google_analytics_v4.db | /data/data/com.forsync/databases/ | ✗ | ✗ | ✗ | ✗ | ✗ | ✗ | ✗ | ✗ | ✗ |
| **17.** | Adobe Creative Cloud | N/A | N/A | ✗ | ✗ | ✗ | ✗ | ✗ | ✗ | ✗ | ✗ | ✗ |
| **18.** | Asus Web Storage Cloud | AWS.db, awsbackup | /data/data/com.ecareme.asuswebstorage/databases/ | ✗ | ✗ | ✗ | ✗ | ✗ | ✗ | ✗ | ✗ | ✗ |
| **19.** | Bitcasa Drive | N/A | N/A | ✗ | ✗ | ✗ | ✗ | ✗ | ✗ | ✗ | ✗ | ✗ |
| **20.** | DropSend Android Cloud | N/A | N/A | ✗ | ✗ | ✓ | ✓ | ✓ | ✓ | ✓ | ✗ | ✗ |
| **21.** | Egnyte | N/A | N/A | ✗ | ✗ | ✗ | ✗ | ✗ | ✗ | ✗ | ✗ | ✗ |
| **22.** | FileBay.co Secure Storage | google_analytics_v4.db | /data/data/com.sflcnetwork.filebayco/databases/ | ✗ | ✗ | ✗ | ✗ | ✗ | ✗ | ✗ | ✗ | ✗ |
| **23.** | Foldersync Lite | foldersync.db | /data/data/dk.tacit.android.foldersync.lite/databases/ | ✓ | ✓ | ✓ | ✓ | ✓ | ✓ | ✓ | ✓ | ✗ |
| **24.** | IDrive Cloud Online Backup | N/A | N/A | ✗ | ✗ | ✗ | ✗ | ✗ | ✗ | ✗ | ✗ | ✗ |
| **25.** | JohoSpace Backup-Restore & Migrate | N/A | N/A | ✗ | ✗ | ✗ | ✗ | ✗ | ✗ | ✗ | ✗ | ✗ |
| **26.** | Mega Cloud | N/A | N/A | ✗ | ✗ | ✗ | ✗ | ✗ | ✗ | ✗ | ✗ | ✗ |
| **27.** | MyCLoud (WD) | wd-files-demo-cache.db, wd-files-cache.db | /data/data/com.wdc.wd2go/databases/ | ✓ | ✓ | ✓ | ✓ | ✓ | ✓ | ✓ | ✓ | ✓ |
| **28.** | Pogoplug Cloud | VISUALS, POGOPLUG | /data/data/com.pogoplug.android/databases/ | ✗ | ✗ | ✗ | ✗ | ✗ | ✗ | ✗ | ✗ | ✗ |
| **29.** | Tesorit CLoud | N/A | N/A | ✗ | ✗ | ✗ | ✗ | ✗ | ✗ | ✗ | ✗ | ✗ |
| **30.** | Yandex Disk | N/A | N/A | ✗ | ✗ | ✗ | ✗ | ✗ | ✗ | ✗ | ✗ | ✗ |
| **31.** | Google Drive | N/A | N/A | ✗ | ✗ | ✗ | ✗ | ✗ | ✗ | ✗ | ✗ | ✗ |

## Appendix 4 – Network Traffic

| **Applications** | | **Protocols Used** | | | **Relevant Stream Data** | |
|---|---|---|---|---|---|---|
| **App ID** | **App name** | **TCP** | **MDNS** | **DNS** | **MDNS** | **DNS** |
| **1.** | Box | ✓ | ✓ | ✓ | ._googlecast._tcp.local..................._googlecast._tcp.local..<br><br>M-SEARCH * HTTP/1.1<br>Host:239.255.255.250:1900<br>ST:urn:schemas-upnp-org:device:MediaRenderer:1<br>Man:"ssdp:discover"<br><br>MX:3 | NOTIFY * HTTP/1.1<br>LOCATION: http://192.168.201.1:0/<br>HOST: 239.255.255.250:1900<br>SERVER: WINDOWS, UPnP/1.0, MicroStack/1.0.4103<br>NTS: ssdp:alive<br>USN: uuid:a343c163-d6a5-4e47-9ae6-1ed85ce82f5d<br>CACHE-CONTROL: max-age=1800<br><br>NT: uuid:a343c163-d6a5-4e47-9ae6-1ed85ce82f5d |
| **2.** | File Expert | ✓ | ✓ | ✓ | M-SEARCH * HTTP/1.1<br>Host:[FF02::C]:1900<br>ST:urn:schemas-upnp-org:device:MediaRenderer:1<br>Man:"ssdp:discover"<br><br>MX:3 | NOTIFY * HTTP/1.1<br>LOCATION: http://192.168.201.1:0/<br>HOST: 239.255.255.250:1900<br>SERVER: WINDOWS, UPnP/1.0, MicroStack/1.0.4103<br>NTS: ssdp:alive<br>USN: uuid:a343c163-d6a5-4e47-9ae6-1ed85ce82f5d<br>CACHE-CONTROL: max-age=1800<br><br>NT: uuid:a343c163-d6a5-4e47-9ae6-1ed85ce82f5d |
| **3.** | GCloud | ✓ | ✗ | ✓ | M-SEARCH * HTTP/1.1<br>Host:239.255.255.250:1900<br>ST:urn:schemas-upnp-org:device:MediaRenderer:1<br>Man:"ssdp:discover"<br><br>MX:3 | NOTIFY * HTTP/1.1<br>LOCATION: http://192.168.201.1:0/<br>HOST: 239.255.255.250:1900<br>SERVER: WINDOWS, UPnP/1.0, MicroStack/1.0.4103<br>NTS: ssdp:alive<br>USN: uuid:a343c163-d6a5-4e47-9ae6-1ed85ce82f5d::upnp:rootdevice<br>CACHE-CONTROL: max-age=1800<br><br>NT: upnp:rootdevice |
| **4.** | Mail.Ru | ✓ | ✗ | ✓ | | (UDP stream): ......cloclo12.datacloudmail.ru....... |
| **5.** | MEO Cloud | ✓ | ✗ | ✓ | | |
| **6.** | Cloud Magic | ✓ | ✗ | ✓ | | |
| **7.** | Drop Box | ✓ | ✓ | ✓ | | |
| **8.** | Degoo | ✓ | ✗ | ✓ | | |
| **9.** | One Drive | ✓ | ✓ | ✓ | (UDP stream): _805741C9._sub._googlecast._tcp.local........... | (UDP stream): c............cid-59DAFF70AD928598.users.storage.live.com..... |
| **10.** | Just Cloud | ✓ | ✗ | ✓ | | |
| **11.** | PCloud | ✓ | ✗ | ✓ | | (UDP stream): ;3...........binapi.pcloud.com..... And........p-par17.pcloud.com..... |
| **12.** | XXL Box Secure | ✓ | ✗ | ✗ | - | - |
| **13.** | ZeroPC | ✓ | ✗ | ✓ | | (UDPstream): m<...........v1.zeropc.com..... |
| **14.** | File Manager | ✓ | ✗ | ✓ | | |
| **15.** | ZipCloud | ✓ | ✗ | ✓ | | (UDP stream): sq..........thumbnails backupgrid.net..... |
| **16.** | 4Sync Cloud | ✓ | ✓ | ✓ | `.............lh5.ggpht.com.....` | `_CC1AD845._sub._googlecast._tcp.local.......................` `_CC1AD845._sub._googlecast._tcp.local.......................` |
| **17.** | Adobe Creative Cloud | ✓ | ✗ | ✓ | `NOTIFY * HTTP/1.1 LOCATION: http://192.168.201.1:0/ HOST: 239.255.255.250:1900 SERVER: WINDOWS,` | `.............cc-api-storage.adobe.io.....` `#............android.clients.google.com.....` |
| | | | | | UPnP/1.0,<br>MicroStack/1.0.4103<br>NTS: ssdp:alive<br>USN: uuid:a343c163-d6a5-4e47-9ae6-1ed85ce82f5d::urn:schemas-upnp-org:device:InternetGatewayDevice:1<br>CACHE-CONTROL: max-age=1800<br>NT: urn:schemas-upnp-org:device:InternetGatewayDevice:1 | |
| **18.** | Asus Web Storage Cloud | ✓ | ✓ | ✓ | .u...........android.clients.google.com..... | M-SEARCH * HTTP/1.1<br>Host:[FF02::C]:1900<br>ST:urn:schemas-upnp-org:device:MediaServer:1<br>Man:"ssdp:discover"<br><br>MX:3 |
| **19.** | Bitcasa Drive | ✓ | ✓ | ✓ | .<br>...........bitcasa.cloudfs.io..... | .............._googlecast._tcp.local..................._googlecast._tcp.local.................._googlecast._tcp.local.................._googlecast._tcp.local.................._googlecast._tcp.local..... |
| **20.** | DropSend Android Cloud | ✓ | ✗ | ✓ | M-SEARCH * HTTP/1.1<br>Host:239.255.255.250:1900<br>ST:urn:schemas-upnp-org:device:MediaServer:1<br>Man:"ssdp:discover"<br>MX:3 | NOTIFY * HTTP/1.1<br>LOCATION: http://192.168.201.1:49860/<br>HOST: 239.255.255.250:1900<br>SERVER: WINDOWS, UPnP/1.0, MicroStack/1.0.4103<br>NTS: ssdp:alive<br>USN: uuid:a343c163-d6a5-4e47-9ae6-1ed85ce82f5d::urn:schemas-upnp-org:service:Layer3Forwarding:1<br>CACHE-CONTROL: max-age=1800<br>NT: urn:schemas-upnp- |
| | | | | | | org:service:Layer3Forwarding:1 |
| **21.** | Egnyte | ✓ | ✗ | ✓ | .>...........decide.mixpanel.com.....<br><br>.............api.mixpanel.com..... | NOTIFY * HTTP/1.1<br>LOCATION: http://192.168.201.1:0/<br>HOST: 239.255.255.250:1900<br>SERVER: WINDOWS, UPnP/1.0, MicroStack/1.0.4103<br>NTS: ssdp:alive<br>USN: uuid:a343c163-d6a5-4e47-9ae6-1ed85ce82f5d::upnp:rootdevice<br>CACHE-CONTROL: max-age=1800<br>NT: upnp:rootdevice |
| **22.** | FileBay.co Secure Storage | ✓ | ✗ | ✓ | 6............www.facebook.com.....<br><br>M-SEARCH * HTTP/1.1<br>Host:239.255.255.250:1900<br>ST:urn:schemas-upnp-org:device:MediaRenderer:1<br>Man:"ssdp:discover"<br>MX:3 | .q...........android.clients.google.com.....<br><br>`H...........r2---sn-aigl6n76.gvt1.com.....<br><br>T............cdn.ywxi.net.....<br><br>NOTIFY * HTTP/1.1<br>LOCATION: http://192.168.201.1:49860/<br>HOST: 239.255.255.250:1900<br>SERVER: WINDOWS, UPnP/1.0, MicroStack/1.0.4103<br>NTS: ssdp:alive<br>USN: uuid:a343c163-d6a5-4e47-9ae6-1ed85ce82f5d_1<br>CACHE-CONTROL: max-age=1800<br>NT: uuid:a343c163-d6a5-4e47-9ae6-1ed85ce82f5d_1 |
| **23.** | Foldersync Lite | ✓ | ✓ | ✓ | ..._googlecast._tcp.local.................._googlecast._tcp.local.<br><br>.............ssl.google-analytics.com.....<br><br>M-SEARCH * HTTP/1.1<br>HOST: 239.255.255.250:1900<br>MAN: "ssdp:discover"<br>MX: 1<br>ST: urn:dial-multiscreen- | NOTIFY * HTTP/1.1<br>LOCATION: http://192.168.201.1:0/<br>HOST: 239.255.255.250:1900<br>SERVER: WINDOWS, UPnP/1.0, MicroStack/1.0.4103<br>NTS: ssdp:alive<br>USN: uuid:a343c163-d6a5-4e47-9ae6-1ed85ce82f5d_1<br>CACHE-CONTROL: max-age=1800<br>NT: uuid:a343c163-d6a5-4e47-9ae6-1ed85ce82f5d_1 |
| | | | | | org:service:dial:1 | |
| **24.** | IDrive Cloud Online Backup | ✓ | ✗ | ✓ | 6............evsweb16 idrivesync.com.....<br><br>.I..........evs idrivesync.com..... | NOTIFY * HTTP/1.1<br>LOCATION: http://192.168.201.1:0/<br>HOST: 239.255.255.250:1900<br>SERVER: WINDOWS, UPnP/1.0, MicroStack/1.0.4103<br>NTS: ssdp:alive<br>USN: uuid:a343c163-d6a5-4e47-9ae6-1ed85ce82f5d_1<br>CACHE-CONTROL: max-age=1800<br>NT: uuid:a343c163-d6a5-4e47-9ae6-1ed85ce82f5d_1 |
| **25.** | JohoSpace Backup – Restore & Migrate | ✓ | ✗ | ✓ | .P..........www.manage .jsbackup.net.....<br><br>M-SEARCH * HTTP/1.1<br>Host:[FF02::C]:1900<br>ST:urn:schemas-upnp-org:device:MediaRenderer:1<br>Man:"ssdp:discover"<br>MX:3 | NOTIFY * HTTP/1.1<br>LOCATION: http://192.168.201.1:0/<br>HOST: 239.255.255.250:1900<br>SERVER: WINDOWS, UPnP/1.0, MicroStack/1.0.4103<br>NTS: ssdp:alive<br>USN: uuid:a343c163-d6a5-4e47-9ae6-1ed85ce82f5d<br>CACHE-CONTROL: max-age=1800<br>NT: uuid:a343c163-d6a5-4e47-9ae6-1ed85ce82f5d |
| **26.** | Mega Cloud | ✓ | ✗ | ✓ | M-SEARCH * HTTP/1.1<br>Host:[FF02::C]:1900<br>ST:urn:schemas-upnp-org:device:MediaRenderer:1<br>Man:"ssdp:discover"<br>MX:3 | NOTIFY * HTTP/1.1<br>LOCATION: http://192.168.201.1:0/<br>HOST: 239.255.255.250:1900<br>SERVER: WINDOWS, UPnP/1.0, MicroStack/1.0.4103<br>NTS: ssdp:alive<br>USN: uuid:a343c163-d6a5-4e47-9ae6-1ed85ce82f5d_1<br>CACHE-CONTROL: max-age=1800<br>NT: uuid:a343c163-d6a5-4e47-9ae6-1ed85ce82f5d_1 |
| **27.** | MyCLoud (WD) | ✓ | ✓ | ✓ | M-SEARCH * HTTP/1.1<br>HOST: 239.255.255.250:1900<br>MAN: "ssdp:discover"<br>MX: 4<br>ST: ssdp:all | NOTIFY * HTTP/1.1<br>LOCATION: http://192.168.201.1:0/<br>HOST: 239.255.255.250:1900<br>SERVER: WINDOWS, UPnP/1.0, MicroStack/1.0.4103<br>NTS: ssdp:alive<br>USN: uuid:a343c163-d6a5-4e47-9ae6-1ed85ce82f5d_1<br>CACHE-CONTROL: max-age=1800<br>NT: uuid:a343c163-d6a5-4e47-9ae6-1ed85ce82f5d_1 |
| **28.** | Pogoplug Cloud | ✓ | ✓ | ✓ | _DC295335._sub._googlecast._tcp.local...... | NOTIFY * HTTP/1.1<br>LOCATION: http://192.168.201.1:0/<br>HOST: 239.255.255.250:1900<br>SERVER: WINDOWS, UPnP/1.0, MicroStack/1.0.4103<br>NTS: ssdp:alive<br>USN: uuid:a343c163-d6a5-4e47-9ae6-1ed85ce82f5d<br>CACHE-CONTROL: max-age=1800<br>NT: uuid:a343c163-d6a5-4e47-9ae6-1ed85ce82f5d |
| **29.** | Tesorit CLoud | ✓ | ✗ | ✓ | .#..........lh3.ggpht.com<br>connectify.....<br><br>.q..........r14---sn-aiglln6k.gvt1.com.....<br><br>M-SEARCH * HTTP/1.1<br>Host:[FF02::C]:1900<br>ST:urn:schemas-upnp-org:device:MediaRenderer:1<br>Man:"ssdp:discover"<br>MX:3 | NOTIFY * HTTP/1.1<br>LOCATION: http://192.168.201.1:0/<br>HOST: 239.255.255.250:1900<br>SERVER: WINDOWS, UPnP/1.0, MicroStack/1.0.4103<br>NTS: ssdp:alive<br>USN: uuid:a343c163-d6a5-4e47-9ae6-1ed85ce82f5d_1<br>CACHE-CONTROL: max-age=1800<br>NT: uuid:a343c163-d6a5-4e47-9ae6-1ed85ce82f5d_1 |
| **30.** | Yandex Disk | ✓ | ✗ | ✓ | M-SEARCH * HTTP/1.1<br>Host:[FF02::C]:1900<br>ST:urn:schemas-upnp-org:device:MediaRenderer:1<br>Man:"ssdp:discover"<br>MX:3 | [&.........<br>uploader7h.disk.yandex.net.....<br>NOTIFY * HTTP/1.1<br>LOCATION: http://192.168.201.1:0/<br>HOST: 239.255.255.250:1900<br>SERVER: WINDOWS, UPnP/1.0, MicroStack/1.0.4103<br>NTS: ssdp:alive<br>USN: uuid:a343c163-d6a5-4e47-9ae6-1ed85ce82f5d<br>CACHE-CONTROL: max-age=1800<br>NT: uuid:a343c163-d6a5-4e47-9ae6-1ed85ce82f5d |
| **31.** | Google Drive | ✓ | ✓ | ✓ | | |